\documentclass[%
 reprint,
superscriptaddress,
 amsmath,amssymb,
 aps,
pra,
]{revtex4-1}

\usepackage{graphicx}% Include figure files
\usepackage{dcolumn}% Align table columns on decimal point
\usepackage{bm}% bold math
\usepackage{textcomp}
\begin{document}

%\preprint{APS/123-QED}

\title{Octave-wide phase-matched four-wave mixing in dispersion engineered crystalline microresonators}% Force line breaks with \\

\author{Shun Fujii}
\affiliation{Department of Electronics and Electrical Engineering, Faculty of Science and Technology, Keio University, Yokohama, 223-8522, Japan}

\author{Shuya Tanaka}
\affiliation{Department of Electronics and Electrical Engineering, Faculty of Science and Technology, Keio University, Yokohama, 223-8522, Japan}

\author{Mika Fuchida}
\affiliation{Department of Electronics and Electrical Engineering, Faculty of Science and Technology, Keio University, Yokohama, 223-8522, Japan}

\author{Hikaru Amano}
\affiliation{Department of System Design Engineering, Faculty of Science and Technology, Keio University, Yokohama, 223-8522, Japan}

\author{Yuka Hayama}
\affiliation{Department of System Design Engineering, Faculty of Science and Technology, Keio University, Yokohama, 223-8522, Japan}

\author{Ryo Suzuki}
\affiliation{Department of Electronics and Electrical Engineering, Faculty of Science and Technology, Keio University, Yokohama, 223-8522, Japan}

\author{Yasuhiro Kakinuma}
\affiliation{Department of System Design Engineering, Faculty of Science and Technology, Keio University, Yokohama, 223-8522, Japan}

\author{Takasumi Tanabe}
\email{takasumi@elec.keio.ac.jp}
\affiliation{Department of Electronics and Electrical Engineering, Faculty of Science and Technology, Keio University, Yokohama, 223-8522, Japan}


%\affiliation{
%\affiliation{
%\affiliation{%

\date{\today}% It is always \today, today,

\begin{abstract}
In this Letter, we report phase-matched four-wave mixing separated by over one-octave in a dispersion engineered crystalline microresonator. Experimental and numerical results presented here confirm that primary sidebands were generated with a frequency shift up to 140~THz, and that secondary sidebands formed a localized comb structure, known as a clustered comb in the vicinity of the primary sidebands. A theoretical analysis of the phase-matching condition validated our experimental observations, and our results good agree well with numerical simulations. These results offer the potential to realize a frequency tunable comb cluster generator operating from 1~\textmu m to mid-infrared wavelengths with a single and compact device.
%\begin{description}
%\end{description}
\end{abstract}

\maketitle

%\tableofcontents

Phase-matched four-wave mixing (FWM) in a whispering gallery mode (WGM) microresonator driven by a continuous wave (CW) laser have been studied for decades~\cite{del2007optical}. In particular, a microresonator frequency comb (Kerr comb) realized via a cascade FWM process makes it possible to achieve broadband optical frequency comb sources characterized as having high repetition rates, compactness and low-energy consumption~\cite{kippenberg2011microresonator}. They provide fascinating applications including precise spectroscopy~\cite{Suh600} and coherent data transmission~\cite{Marin-Palomo2017}. A Kerr comb, and particularly a dissipative Kerr soliton (DKS), which is a stable low noise state, features a coherent broad comb spectrum with a smooth envelope~\cite{herr2014temporal}. However, spectral broadening over one octave remains a challenge mainly because of the limitation imposed by group velocity dispersion (GVD). Spectral broadening utilizing dispersion engineering has been achieved with Cherenkov radiation, which can be understood in terms of the coherent dispersive wave in the frequency domain emitted from a soliton propagating along a resonator~\cite{brasch2016photonic}. A dispersive wave induces asymmetrical spectral profiles and occurs at the point where a simple phase-matching condition is satisfied, and it is determined by the signs and values of higher-order (i.e., third-, fourth- and fifth-order) dispersion parameters.

Higher-order dispersion plays important roles not only as regards a Kerr soliton in an anomalous GVD regime, but also in a weak normal GVD regime. Since a normal dispersion usually does not allow modulation instability (MI) near the pump, phase-matched FWM may be considered to occur only in an anomalous dispersion resonator. However, higher-order dispersion, particularly even orders of dispersion, enables a resonant MI process, called clustered frequency combs, to occur far from the pump mode~\cite{matsko2016clustered,Fujii2017,Sayson:18}. Clustered combs, characterized by FWM generation with parametric sidebands that have a large-frequency shift, have been reported using an $\mathrm{MgF_2}$ microresonator~\cite{matsko2016clustered,Sayson:18} and silica microtoroids~\cite{Fujii2017}. The fact that clustered combs have the potential to utilize microcomb source emitting in the 1.0 to 3.5~\textmu m wavelength region with just a near-infrared pump indicates interesting potential applications, for example laser processing and optical communication. In particular, T-band (1.0--1.26~\textmu m) and O-band (1.26--1.36~\textmu m) communication has recently attracted attention due to a rapid increase in data traffic, especially for use in short distance communication such as local area networks~\cite{823491}. In addition, optical parametric oscillation towards mid-infrared wavelengths has remained important for many applications including spectroscopy. The challenge of tunable mid-infrared generation via optical parametric amplification has been met with a silicon nitride waveguide, however it requires a conventional pulse laser to provide the pump and signal~\cite{Kowligy:18}. 

A pure parametric oscillation (i.e., signal and idler pair)  unaccompanied by a clustered comb structure also offers compact, low-cost and widely tunable light sources or a non-classical photon-pair source for quantum applications.  In both cases, the oscillation wavelengths are essential and interesting subjects for investigation as regards clustered frequency comb and parametric sideband generation. Thus far, the maximum frequency shift observed experimentally is 85~THz (from 1207~nm to 1930~nm) in a silica microresonator via a near-infrared pump~\cite{Sayson:17}, and it is considered that the wavelength range is limited by the strong absorption of fused silica at longer wavelengths. Therefore, the use of crystalline resonators instead of silica resonators is straightforward because crystalline materials are transparent in the visible to mid-infrared wavelength region. One further challenge is to control the oscillation wavelength with respect to dispersion engineering in addition to pump wavelength tuning.

In this Letter, we describe our experimental demonstration of a large frequency shift FWM that exceeds one octave by using a 1.55~\textmu m pump in a dispersion engineered  $\mathrm{MgF_2}$ crystalline microresonator. The signal and idler oscillate at 1137~nm and 2433~nm, respectively, which is the largest frequency shift (up to 140~THz) achieved in phase-matched FWM. The resonance mode is identified with a precise dispersion measurement method, and the experimental results are in good agreement with calculations.

\begin{figure}[!t]
	\centering
	\includegraphics[]{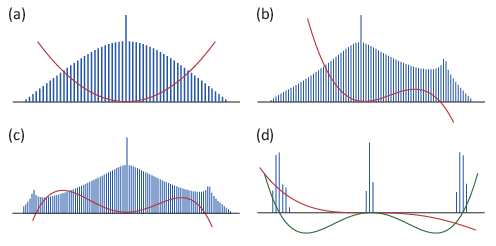}
	\caption{Kerr comb spectrum and corresponding integrated dispersion (solid red line). Dissipative Kerr solitons dominated by $D_2$ (a), with a single peak (b) and double peaks (c) affected by $D_3$ and $D_4$, respectively. Clustered comb formed in a normal dispersion regime from the balance between $D_2$ and $D_4$. The green solid line shows phase-matching points.}
	\label{fig1}
\end{figure}

Microresonator dispersion can be expressed with respect to resonance frequency around the center frequency $\omega_0$ as the Taylor expansion: $\omega_\mu=\omega_0 + \sum(D_j \mu^j)/(j!) (j\geq1)$, where $\mu$ is the relative mode number, $D_1/2\pi$ is the free-spectral range (FSR) of the resonator, $D_2$ is the second-order dispersion related to GVD, and $D_3$, $D_4$, $\cdots$ are the higher-order dispersion. The above expression is often used to represent the integrated dispersion as a function of $\mu$, $D_\mathrm{int} = (1/2)D_2\mu^2 + (1/6)D_3\mu^3 + (1/24)D_4\mu^4$ when omitting the above fifth-order dispersion. In this work, we only took third- and fourth-order dispersion into account since this is sufficient to validate our experimental results presented here. Note that there has been a numerical study that accounts for the effect of higher-order dispersion (e.g., fifth- and sixth-order) on the spectrum of a Kerr comb~\cite{Bao:17}. Figure~\ref{fig1} shows schematics of a Kerr comb spectrum and corresponding dispersion $D_\mathrm{int}$ (solid red line). DKS forms a $\mathrm{sech}^2$-shaped envelope when $D_2$ is dominant [Fig.~\ref{fig1}(a)], however DKS with dispersive waves with single [Fig.~\ref{fig1}(b)] and double peaks [Fig.~\ref{fig1}(c)] are observed when $D_3$ and $D_4$ appear, respectively. A clustered comb is shown in Fig.~\ref{fig1}(d), where the sign of $D_2$ is negative (normal dispersion), and that of $D_4$ is positive. The phase-matched wavelength for initial parametric sidebands can be estimated simply using frequency difference value defined as 
($\omega_{0+\mu}+\omega_{0-\mu}-2\omega_0)=D_2 \mu^2 +(1/12)D_4 \mu^4$ (solid green line). Here, $D_\mathrm{int}$ shows the cubic function affected by $D_3$, however only even orders of dispersion play a role as regards the phase-matching condition for the initiation of sidebands. The formation scheme for a clustered comb is introduced in \cite{Sayson:18} as follows; after primary sideband (signal and idler pair) generation, secondary sidebands are generated via degenerate FWM from one primary sideband experiencing anomalous dispersion, which is ordinarily low frequency (longer wavelength) signal light. Once the secondary sidebands have been generated, non-degenerate FWM will occur and form other comb lines in the vicinity of the idler and pump lights.

\begin{figure}[!t]
	\centering
	\includegraphics[]{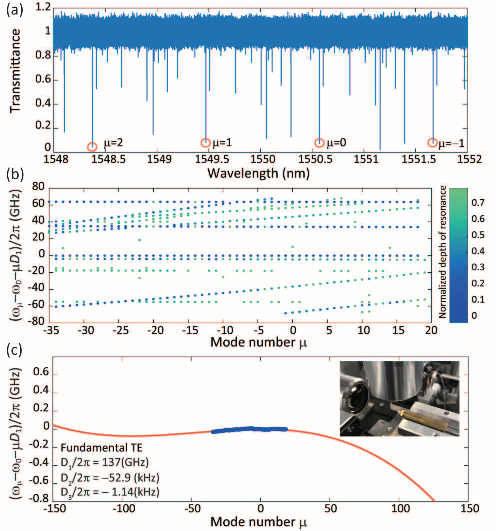}
	\caption{(a) Transmission spectrum of fabricated $\mathrm{MgF_2}$ microresonator. Red circles indicate resonances belonging to same mode family that we used as pump mode in comb experiment. (b) Experimentally observed mode structure, where dots show the position of resonances, in a wavelength from 1530 to 1590~nm, limited by bandwidth of fiber comb source.  (c) The calculated dispersion and measured result of 55 resonances for fundamental TE mode. Insets shows picture of experimental setup for tapered fiber coupling.}
	\label{fig2}
\end{figure}

In our experiment, we used an $\mathrm{MgF_2}$ crystalline microresonator with a diameter of 508~\textmu m and a curvature of 36~\textmu m, corresponding to a 137~GHz FSR, fabricated by computer-controlled machining. The fabrication is described in detail elsewhere~\cite{cleofujii}. The structure and corresponding dispersion of the resonator were numerically studied in advance to realize octave-wide phase-matched FWM. We coupled a laser light into the resonator by using a tapered optical silica fiber. First, we performed a highly accurate dispersion measurement on the fabricated resonator assisted by a fiber laser comb~\cite{del2009frequency}. Figure~\ref{fig2}(a) shows the transmission spectrum around 1550~nm measured with a broadly tunable external cavity diode laser. There are different mode families in the fundamental mode FSR that are indicated as red circles. The measured quality factor of $\mu=0$ was $8\times10^7$. The measured mode structure is shown in Fig.~\ref{fig2}(b), which enables us to identify the fundamental mode (pump mode). We calculated the  resonator dispersion with the finite element method (COMSOL Multiphysics), and obtained the  theoretical dispersion of the fundamental TE mode shown in Fig.~\ref{fig2}(c) (solid red line). The measured dispersion (blue dots) agrees well with the calculated dispersion $(\omega_\mu - \omega_0 - D_1 \mu)/2\pi$, which confirmed that our ultraprecise lathe enables us to fabricate a calculated (pre-designed) cross-sectional shape. From this precise resonator dispersion measurement, we identified the resonant mode with a weak normal dispersion, which is required for optical parametric sideband generation far from the pump.

\begin{figure}[!t]
	\centering
	\includegraphics[]{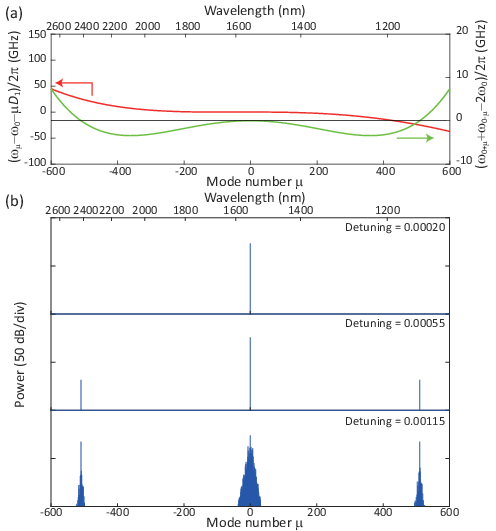}
	\caption{(a) Calculated integrated dispersion $D_{\mathrm{int}}$ (solid red line) and frequency difference value $\omega_{0+\mu}+\omega_{0-\mu}-2\omega_0)$ (solid green line) of fundamental TE mode. The points at which the green line crosses black reference line predict the phase-matched wavelength. (b) Simulation results with increasing pump detuning. The parameters are as follows; $Q_{\mathrm{int}}=1\times10^8$, $Q_{\mathrm{ext}}=4\times10^8$, $g=0.004$, and $P_{\mathrm{in}}=300$~mW.}
	\label{fig3}
\end{figure}

Next, we numerically investigate the possibility of clustered comb generation. To obtain the parametric oscillation, the bandwidth of the parametric gain envelope must overlap the resonant modes in the resonator. Parametric gain bandwidth has been analytically studied~\cite{Sayson:17}, and it was shown that the gain bandwidth tends to decrease as the frequency shift becomes larger. If the gain bandwidth falls below 1-FSR of the resonator, the possibility of phase-matched FWM also significantly decreases. We conducted a numerical simulation using the Lugiato-Lefever equation (LLE) to confirm both the phase-matched wavelength and the effective parametric gain described above. The LLE including higher-order dispersion $D_k (k\geq2)$ is expressed as follows~\cite{PhysRevA.87.053852}:

\begin{figure}[!t]
	\centering
	\includegraphics[]{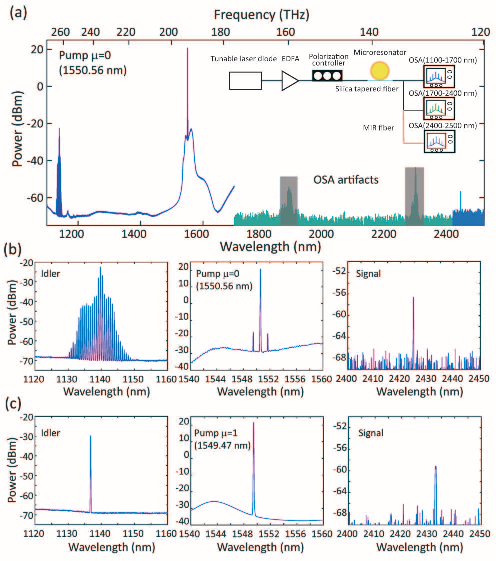}
	\caption{(a) Experimentally observed spectrum at a pump mode $\mu=0$ when using different OSAs (depicted in different colors). The inset shows the experimental setup. MIR fiber was used above a 2.4~\textmu m wavelength to avoid unwanted attenuation. (b) Detailed spectra of interest, which show a clustered comb formation in the vicinity of 1140~nm and the pump wavelength, and that also confirmed the signal wavelength at 2425~nm. (c) Observed spectra when pumping in an adjacent mode $\mu=1$, indicating an FWM pair of 1137 and 2433~nm.}
	\label{fig4}
\end{figure}

\begin{equation}
\begin{split}
\frac{\partial A(\phi,t)}{\partial t} = -\left( \frac{\kappa_{\mathrm{tot}}}{2} +i \delta_0 \right)A - i \sum_{k=2} \frac{D_k}{k!} \left( \frac{\partial}{i \partial \phi} \right) ^k A \\
+ i g |A|^2 A + \sqrt{\kappa_{\mathrm{ext}}}A_{\mathrm{in}}, \label{Eq.1}
\end{split}
\end{equation}
where $\phi $ is the azimuthal angle along the circumference of the resonator, $t$ is the time describing the evolution of the field envelope, $\kappa_\mathrm{tot}$ is the total decay rate given by the sum of the intrinsic loss $\kappa_\mathrm{int}$ and the external coupling rate $\kappa_\mathrm{ext}$, $\delta_0$ is the pump detuning from the resonance, $g$ is a nonlinear coefficient, and $A_{\mathrm{in}}=\sqrt{P_{\mathrm{in}}/\hbar \omega_p}$ is the input field. It should be noted that $\kappa$ is the related to the quality factor ($Q$) in $Q=\omega_0/\kappa$. The dispersion used in the LLE simulation is shown in Fig.~\ref{fig3}(a), where solid red line and solid green lines represent the integrated dispersion and the frequency difference value, respectively (See also Fig.~\ref{fig1}). The quality factor in this simulation follows our experimental result, and other parameters are obtained by FEM calculation and from physical values.  From the zero-cross point of the solid green line, we estimate the position of the initial sidebands via degenerate-FWM, which is approximately at mode number $\mu = \pm 510$. The values correspond to a the frequency shift of 70~THz from the pump frequency, and they nearly reach the mid-infrared wavelength region with 1.55~\textmu m pumping.
Figure~\ref{fig3}(b) shows the simulated spectra at different normalized detuning values $t_R \delta_0$ where $t_R$ is the round-trip time. When the detuning reached the threshold, primary sidebands  were generated at the theoretically predicted wavelength, and they finally became three clustered combs as we expected.

With the numerical simulation in mind, we performed an experimental demonstration in an $\mathrm{MgF_2}$ microresonator. The input CW laser was amplified using an erbium-doped fiber amplifier (EDFA), and the polarization was adjusted using a polarization controller. We observed the comb spectrum with three optical spectrum analyzers (OSA) to cover the octave-wide wavelength, and by splitting the output light with an optical coupler [see inset of Fig.~\ref{fig4}(a)]. First, we pumped the resonant mode $\mu=0$ (1550.56~nm) with an input power of 350~mW. With careful tuning of the pump wavelength, we observed stable optical parametric sidebands, where the wavelengths were 1140 and 2425~nm, respectively [Fig.~\ref{fig4}(a)].  Figure~\ref{fig4}(b) is a magnification of a significant region, and it confirms that high frequency sidebands formed a clustered comb with a 1-FSR spacing. Unfortunately, the secondary sidebands in the vicinity of the pump line were concealed by the amplifier noise of the EDFA, however we measured the single FSR sidebands on both sides of the pump. Next, we tuned the laser wavelength to another resonant mode $\mu=1$ (1549.47~nm), and also observed the initial sidebands with wavelengths of 1137 and 2433~nm [Fig.~\ref{fig4}(c)]. These experimental observations agree well with our simulation results, and the frequency shift of the FWM sidebands exceeds one octave, which is the largest frequency difference (up to 140~THz).

\begin{figure}[!t]
	\centering
	\includegraphics[]{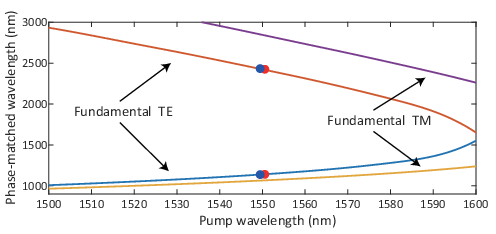}
	\caption{Phase-matched wavelength of fundamental mode as a function of the pump wavelength in $\mathrm{MgF_2}$ used in this work. Experimental plots are shown as blue and red circles. }
	\label{fig5}
\end{figure}

Nevertheless, the power of the low-frequency sidebands ranges from very weak to barely measurable compared with that on the pump and high-frequency side.   Here, we consider the reason for this power asymmetry as follows: the most probable reason is the strong material absorption of silica above 1.9~\textmu m. Although the fluoride material used as a resonator is transparent even at mid-infrared wavelengths, the tapered fiber and optical coupler used in this experiment were made of silica. The second reason is the phase-mismatch between the tapered fiber and the optical mode since the waist of the tapered fiber is optimized for a 1.55~\textmu m pump wavelength. Therefore, the coupling efficiency is low at wavelengths  far from the pump mode.
A simple way to overcome the problem is to utilize low-OH silica fiber as a tapered coupler as in a previous work that reported Kerr comb generation at a wavelength of 2.45~\textmu m ~\cite{wang2013mid}. However, the tapered region should be optimized for the pump wavelength even with this method. Another solution, although it demands a more complex setup, would be to employ an add-drop system, where we must use a silica tapered fiber for the pump (add side) and a waist optimized low-OH or chalcogenide (ChG) tapered fiber for low-frequency light (drop side), thus realizing high coupling efficiency. In particular, ChG tapered fiber is more suitable for the mid-infrared region above 2~\textmu m~~\cite{Lecaplain2016}.

Figure~\ref{fig5} shows the theoretically predicted phase-matched wavelength when changing pump wavelength and the experimentally measured wavelengths (blue and red circles). As a previous work demonstrated, the wavelength is tunable by employing a clustered comb scheme, and our results also follow the theoretical prediction. In our experiment, we limited the pump wavelength to only around 1.55~\textmu m because wavelengths below 1.55~\textmu m are outside  wavelength range of the narrow-linewidth pump laser that we used.

In conclusion, we have demonstrated octave-wide phase-matched FWM in a crystalline optical microresonator, which we designed to realize clustered comb generation via octave-wide separated FWM. Although identification of the resonant mode is generally hard because it is close to zero dispersion, the employed precise dispersion measurement allowed us to distinguish mode families, and to pump the desired mode. The analysis of phase-matching condition including higher-order dispersion and numerical simulation supported our experimental observation of clustered comb formation in both 1.1 and 2.4~\textmu m wavelength, which could promise the localized microcomb formation in an unexplored wavelength region, for example, the T-band (1.0--1.26~\textmu m), O-band (1.26--1.36~\textmu m) and mid-infrared region, from compact Kerr nonlinear devices.

\section*{Funding.}
Japan Society for the Promotion of Science (JSPS) KAKENHI under Grant Number JP18J21797 and Grant-in-Aid for JSPS Fellow; Amada Foundation; The Ministry of Education, Culture, Sports, Science and Technology (MEXT) Q-LEAP.


% Bibliography
%\bibliography{clustered_comb}

\end{document}